\documentclass[multphys,vecphys]{svmult}

\usepackage{makeidx}         
\usepackage{graphicx}        
\usepackage{multicol}        
\usepackage[bottom]{footmisc}

\makeindex             
\newcommand{\lmax}{\mbox{$L_{\rm max}$}}
\newcommand{\mmax}{\mbox{$M_{\rm max}$}}
\newcommand{\mmaxphys}{\mbox{$M_{\rm max}^{\rm phys}$}}
\newcommand{\mmin}{\mbox{$M_{\rm min}$}}
\newcommand{\dr}{\mbox{${\rm d}$}}
\newcommand{\msun}{\mbox{$M_{\odot}$}}
\newcommand{\pint}{\mbox{$P_{\rm int}$}}

\begin{document}

\title*{The Maximum Mass of Star Clusters}
\author{M. Gieles\inst{1}\and
S.S. Larsen\inst{1,2}\and
M.R. Haas\inst{1}\and
R.A. Scheepmaker\inst{1}\and
N. Bastian\inst{3}}
\institute{Utrecht University 
\texttt{gieles@astro.uu.nl}
\and European Southern Observatory (Garching)
\texttt{larsen@astro.uu.nl}
\and University College London
\texttt{bastian@star.ucl.ac.uk}
}
%
%
\maketitle

%
\vspace{-0.1cm}
\section{A Maximum from Size-of-Sample Effects or Physics?}
\label{sec:1}
If an universal untruncated cluster initial mass function (CIMF) of the
form $N(M)\dr M = C\,M^{-2}\dr M$ is assumed, the mass of the most
massive star cluster  in a galaxy ($\mmax$) is the result of the
size-of-sample (SoS) effect. This implies a dependence of \mmax\ on
the total number of clusters ($N$). For a power-law index of -2, the
constant $C=\mmax$ and $N$ follows from integrating the CIMF from
$\mmin$ to $\mmax$, resulting in $N=\mmax/\mmin$. Since the cluster
luminosity function (CLF) is also a power-law distribution, with a
comparable index, a similar relation holds for the luminosity of the
brightest cluster in a galaxy ($\lmax$) and $N$, which has been
observed \cite{2003dhst.symp..153W,larsen02}. An attempt to
compare $\mmax$ in a sample of galaxies with the star formation rate
(SFR) has shown a similar relation \cite{weidner04}. However, finding
the most massive cluster in a galaxy is not trivial, since star
clusters fade rapidly due to stellar evolution. For example, a 1 Gyr
old cluster of $10^6\,\msun$ has about the same luminosity as a 4 Myr
old cluster of $10^4\,\msun$.  The SoS effect also implies that \mmax\
within a cluster population increases with equal logarithmic intervals of
age. This is because the number of clusters formed in logarithmic age
intervals increases (assuming a constant cluster formation rate). This
effect has been observed in the SMC and LMC \cite{hunter03}. The
observations of this increase argues for a $\mmax$ (in the
LMC and SMC) that is determined by sampling statistics, {\it or} a
physical upper limit that is higher than the $\mmax$ following from
statistics.

Based on the maximum pressure ($\pint$) inside molecular clouds, it
has been suggested that a physical maximum mass ($\mmaxphys$) should
exist, which scales as $\mmaxphys \propto \pint^{1/2}$
\cite{elmegreen01}. The ISM pressure in a galaxy scales approximately
as the square of the column density of molecular gas ($\Sigma^2$), and
when assuming that \pint\ is determined by the ISM pressure
(i.e. pressure equilibrium), then $\mmaxphys\propto\Sigma$. Since the
star formation rate (SFR) scales in another way with $\Sigma$, namely
SFR$\,\propto\Sigma^{1.4}$, and since \mmaxphys\ is independent of the
size of the galaxy ($A$), for a certain minimum $A$ and SFR a
$\Sigma_{\rm crit}$ should exist where $\mmax=\mmaxphys$. For galaxies
where $\Sigma>\Sigma_{\rm crit}$, $\mmaxphys$ is lower than the
$\mmax$ determined by sampling statistics. To observe signatures of the presence of \mmaxphys, one should
look in big galaxies where $\Sigma$ (or the SFR) is high.

\begin{figure}
\vspace{-0.cm}
\centering
\includegraphics[height=2.5cm]{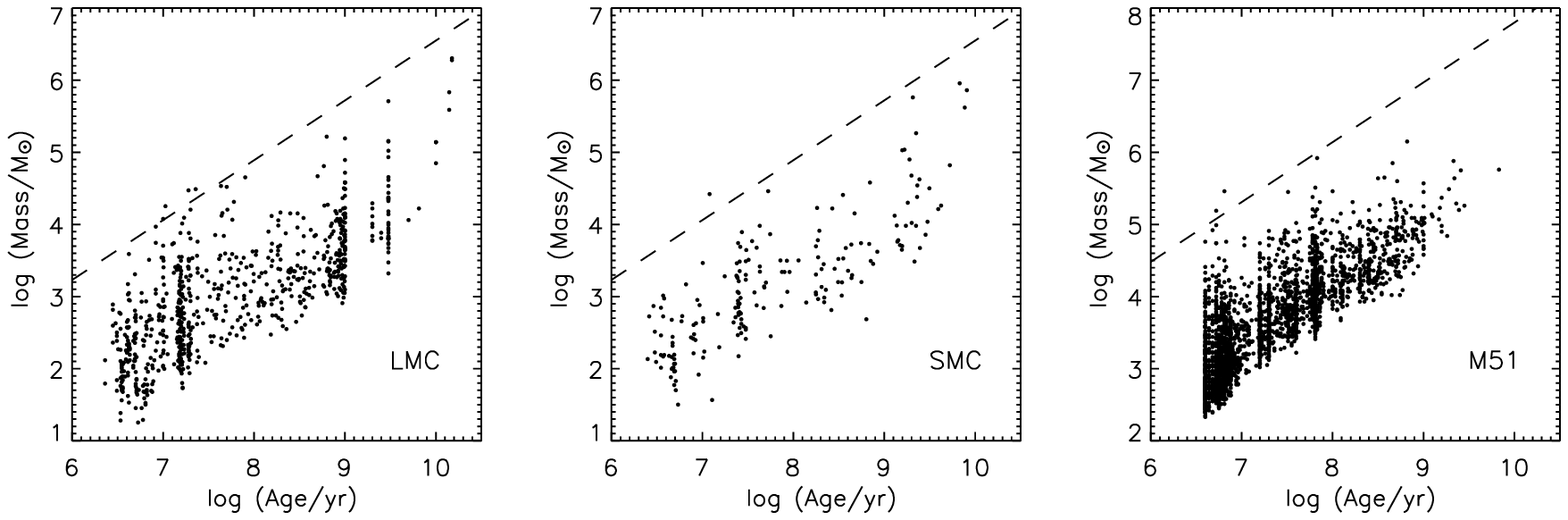}
\includegraphics[height=2.5cm]{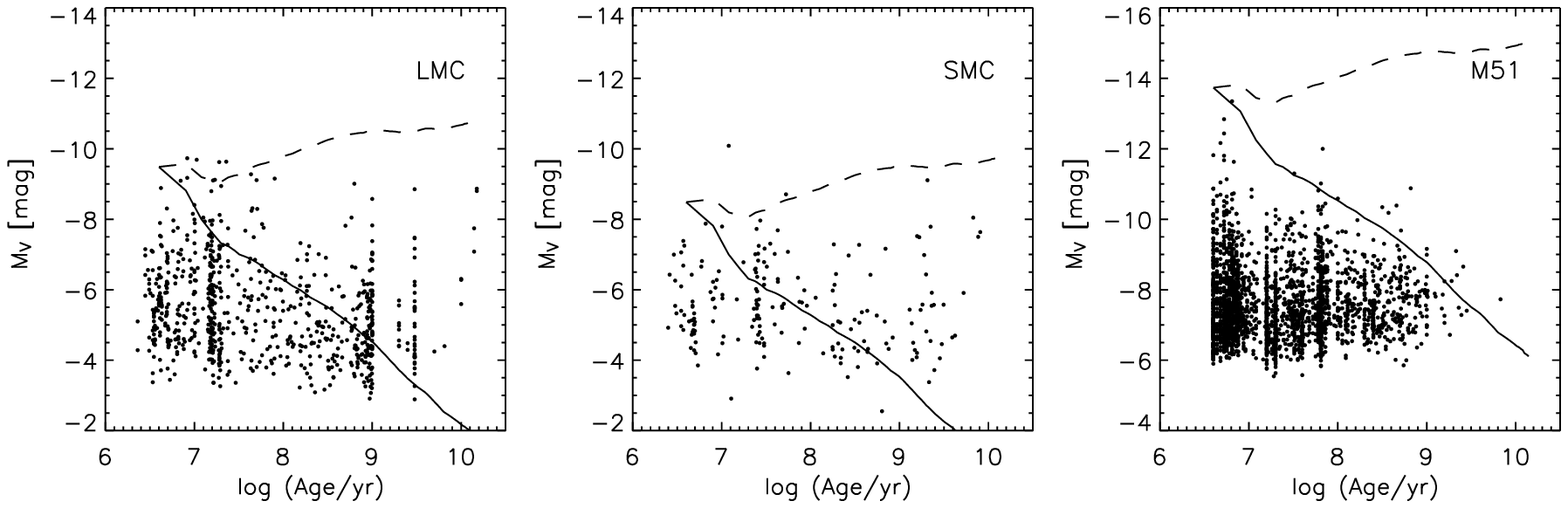}
\caption{{\it Top:} Age-mass diagrams of clusters in the LMC, SMC and
M51. {\it Bottom:} Age-luminosity diagrams for the same clusters. The
SoS relations are shown as dashed lines. Fading lines from SSP models
are shown as full lines (from \cite{gieles06a}).}
\label{fig:1}       
\end{figure}

\vspace{-0.2cm}
\section{The Size-of-Sample Effect in M51}
\label{sec:2}

A good candidate galaxy, which is big and has a high SFR, is M51. We
used the 1052 star clusters identified by \cite{bastian05} to study
the SoS relation of $\mmax$ with log(age). In Fig.~\ref{fig:1} we show
a comparison between the clusters in the LMC (left), the SMC (middle)
and M51 (right). In the top panel we compare the age-mass diagrams,
where we have overplotted the predicted SoS increase of $\mmax$ with
log(age) as dashed lines, based on a power-law CIMF with index
-2. As was shown by \cite{hunter03}, $\mmax$ in the LMC and SMC
follows this prediction quite well. In M51, however, there is a lack
of old ($>\sim\!\!10^8\,$yr), massive ($>\sim\!\!10^6\,\msun$) clusters. In
the bottom panels we show the luminosity (magnitude) {\it vs.}
log(age). The SoS relation for $\mmax$ is converted to $\lmax$ using
the GALEV SSP models \cite{anders03} and is almost a horizontal
line. Fading lines, scaled to the brightest clusters at young ages,
are shown as full lines. The brightest cluster {\it vs.} log(age) in
M51 follow this fading line of a $5\times10^5\,\msun$ cluster quite
well, similar to what was found for the ``Antennae'' galaxies (for a
$10^6\,\msun$ cluster) \cite{zhang99}. {\it This suggests that the
cluster mass function in M51 and the ``Antennae'' galaxies is
truncated around $\sim0.5$-$1.0\times10^6\,\msun$.}

\vspace{-0.2cm}
\section{The Integrated Star Cluster Luminosity Function}
\begin{figure}[!t]
\centering
\includegraphics[height=2.1cm]{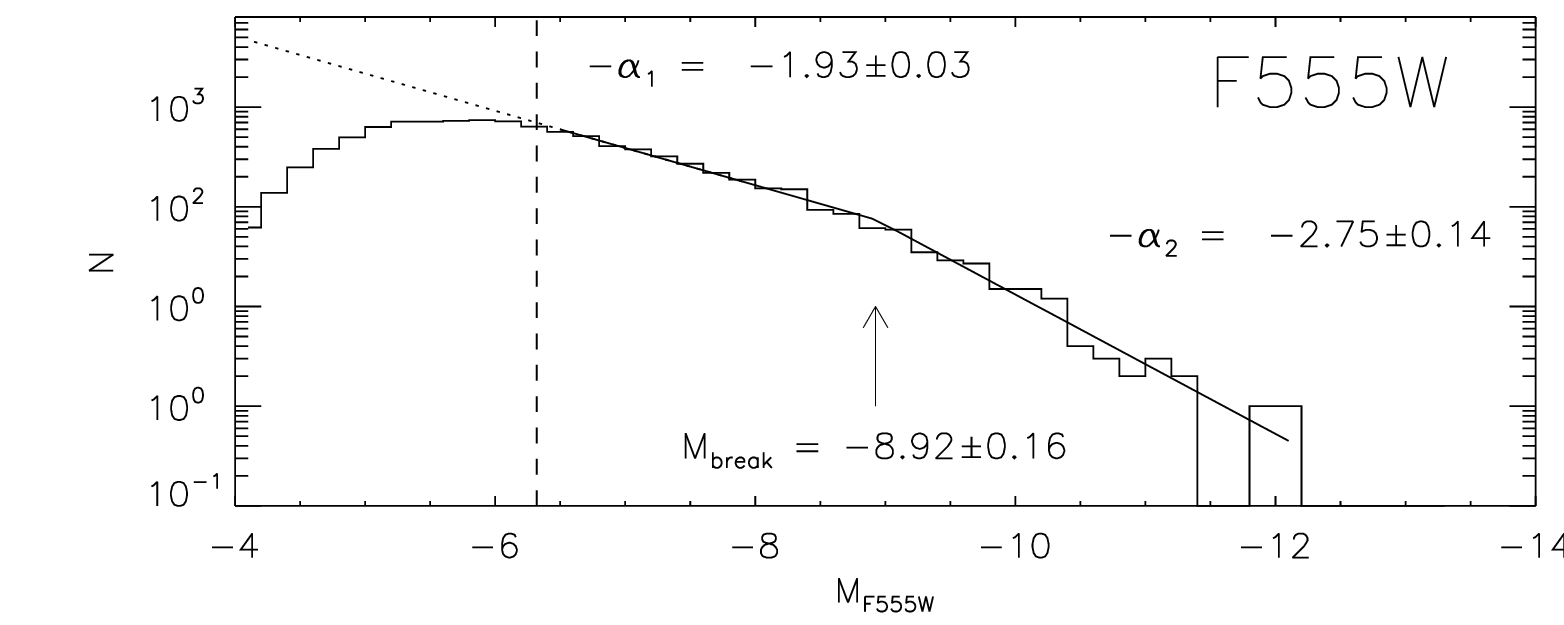}\includegraphics[height=2.1cm]{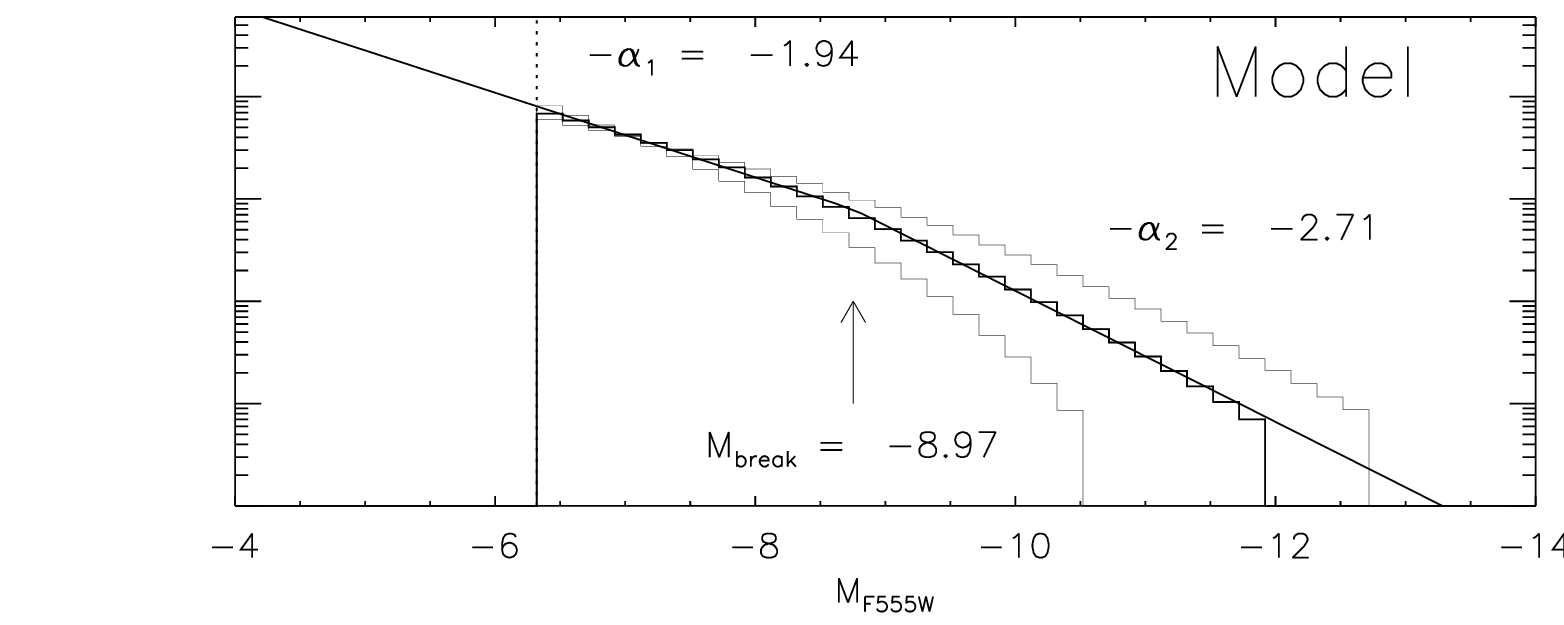}

\caption{{\it Left:} Observed CLF of $\sim6000$ star clusters in M51. {\it
Right:} Modelled CLF of a population with
$\mmaxphys=5\times10^5\,\msun$. (From \cite{gieles06b}).}
\label{fig:2}       
\end{figure}
Since the age determination from broad-band colours has limitations, we
want to have an independent check of the {\it truncated mass function
scenario}, without relying on age determination. Therefore, we model
for two scenarios the integrated cluster luminosity function (CLF) of
a population which has formed with a constant cluster
formation rate (CFR): 1.) $\mmax$ is determined by SoS effects and
increases with log(age) and 2.)  $\mmax=\mmaxphys$ is constant with
log(age). The CLF in case 1.) is a power-law distribution, with an
index similar to the underlying mass function. This has been observed
for various spiral galaxies and the LMC and SMC
\cite{gieles06a,larsen02}. The resulting CLF of scenario 2.) is better
described by a double power-law distribution, for which the location of
the break is determined by $\mmaxphys$. On the bright side of the CLF the
index is smaller than $-2$ (i.e. steeper), and on the faint side it is
$\simeq-2$. The steeper bright side is because a truncation in the mass
function will be spread out over a range of luminosities due to the
age spread in the population and fading of clusters in time
(e.g. young clusters with $\mmaxphys$ are brighter than old clusters with
$\mmaxphys$). Tentative evidence for a double power-law CLF was observed
for NGC~6946 and M51 \cite{gieles06a}.

Recently, the Hubble Heritage project released new {\it HST/ACS} data
of M51, covering the entire disc with 6 pointings. We used this
dataset and selected clusters based on the size. All sources found
with SExtractor ($\sim~\!\!\!70\,000$), were compared to (extended)
cluster profiles convolved with the camera PSF. Around $6\,000$
sources, above a conservative completeness limit, were found to be
more extended than the instrumental PSF. The resulting CLF of this
sample shows a pronounced double power-law behaviour and is very
similar to what was found from the models (see Fig.~\ref{fig:2}).

Several predictions from the CLF model are found back in the
observations: 1.) The power-law index on the bright side ($-\alpha_2$)
increases when going to bluer filters. This is because clusters fade
more rapidly in the bluer filters, which spreads out the luminosity of
$\mmaxphys$ over a larger range of magnitudes; 2.) The break in the
CLF shifts to brighter luminosities when going to redder filters. This
is because the majority of the clusters with the break luminosity is
red (see \cite{gieles06a} for details). The best agreement between
data and model, taking into account cluster disruption and extinction,
is for $\mmaxphys=5\times10^5\,\msun$. A similar double power-law CLF
was observed for the ``Antennae'' clusters \cite{whitmore99}, although
with a break $1.4$ mag brighter, implying that $\mmaxphys({\rm
Antennae})=4\times\mmaxphys({\rm M51})\simeq2\times10^6\,\msun$. We
note that a direct comparison between the CLF of ``Antennae'' clusters
and the one following from our model is dangerous because of the
non-constant CFR in the ``Antennae'' galaxies. Nevertheless, {\it the
observed break in the CLF is an independent confirmation of the
truncated mass function scenario, confirming the results from the SoS
comparison of Sect.~\ref{sec:2}.}
\vspace{-0.5cm}
\section{The Environmental Dependency of $\mmaxphys$ }
The difference between $\mmax$ in the ``Antennae'' galaxies and in M51
and the recently discovered super-massive star clusters
\cite{bastian06} (also Bastian in these proceedings), suggest an
environmental dependent $\mmaxphys$.  We looked for variations of the
bend location {\it within} M51 at different galactocentric radii
($R$). If $\mmax\propto\Sigma$, and $\Sigma\propto\exp(-R/R_{\rm h})$,
then $\mmax \propto \exp(-R/R_{\rm h})$, with $R_{\rm h}$ the disc
scale length of molecular gas. We found a correlation, since in three
radial bins with $\bar{R}/{\rm kpc}=[1.5,\,4.5,\,7]$ we find the bend
at $M_V=[-8.6,-8.5,-7.7]$ \cite{haas06}. Although the errors in the
fit are large ($\pm 0.2\,$mag), the decreasing \mmaxphys\ with $R$ is
a third argument supporting the truncated mass function scenario in
M51.
\vspace{-0.6cm}
\section{Final Thoughts}
Our observations of a truncation of the {\it integrated} mass function
does not necessarily imply that a truncation is visible in the
CIMF, since there $N$ is much lower. Therefore, the observations of an
untruncated CIMF in M51 \cite{bik03} and the ``Antennae'' galaxies
\cite{zhang99} are not in disagreement with what we discuss here. In
addition, the scaling of $\lmax$ with $N$ is expected to be determined
by the SoS effect, since the brightest cluster is generally young
($<\,$10 Myr). The number of clusters in a young sample is too small
to sample the mass function up to \mmaxphys.

\begin{acknowledgement}
I thank Bruce Elmegreen for interesting discussions during the meeting
in Concepci\'{o}n and the organisers for a great conference and a nice
asado!
\end{acknowledgement}

%
%
%
%
%

%
%



\printindex
\end{document}